# Title: Motor Learning Mechanism on the Neuron Scale

**Authors:** Peilei Liu[1]*, Ting Wang[1]

**Affiliations:**

[1]College of Computer, National University of Defense Technology, 410073 Changsha, Hunan, China.

*Correspondence to: plliu@nudt.edu.cn

**Abstract**: Based on existing data, we wish to put forward a biological model of motor system on the neuron scale. Then we indicate its implications in statistics and learning. Specifically, neuron's firing frequency and synaptic strength are probability estimates in essence. And the lateral inhibition also has statistical implications. From the standpoint of learning, dendritic competition through retrograde messengers is the foundation of conditional reflex and "grandmother cell" coding. And they are the kernel mechanisms of motor learning and sensory-motor integration respectively. Finally, we compare motor system with sensory system. In short, we would like to bridge the gap between molecule evidences and computational models.

**Main Text:** Great strides have been made in the research of motor learning (*1-5*). Until now however, there still exists a gap between existing models and physiological data (*6-8*). Control models such as the internal forward models (*1, 2*) and optimal control models (*3*) are mainly on the module scale. Learning models such as CMAC are difficult for biological implementation (*4, 5*). Based on existing data and theories, we wish to put forward a quantitative motor learning model on the neuron scale. Inspired by the "self-organization" idea (*9*), we only make local rules about neuron and synapse, and the neural network will emerge automatically. Moreover, both excitory and inhibitory neurons share the same framework, merely different in details. Motor neurons in this paper include those in the cerebellum, DCN (deep cerebellar nuclei) and basal ganglia, but excluding Pyramid cells in the cerebral motor area. All $c_i$ in this paper are constants, and they have different meanings in different paragraphs. Information about sensory memory can be found in our previous work (*10*).

Motor neuron model is as follows (**N**): N1) $f = c_1(1-e^{-c_2 h|\sigma|})$, $\sigma = \sum_{i=1}^{n} f_i$, $f_i = a_i w_i g_i$ for excitory synapse, and for inhibitory synapse $f_i = (c_3 - a_i) w_i g_i$ ($a_i < c_3$), where f is the soma's spike frequency or hyperpolarized degree, h is hormone factor, $f_i$ is dendritic spike frequency or hyperpolarized degree, $a_i$ is sensitivity of membrane channels, $w_{ij}$ is the synaptic strength, $g_i$ is the presynaptic spike frequency. N1 is inspired by the MP model (*11*) and Hodgkin-Huxley model (*12*). N2) $\frac{da_i}{dt} = -c_4 g_i a_i$ when depolarized, and $\frac{da_i}{dt} = c_5(g_i + c_6)(c_3 - a_i)$ when hyperpolarized. Generally speaking, membrane channels become fatigued when a dendrite is depolarized. And they will recover gradually when hyperpolarized (see Fig. 1). In reality, a dendrite has many synapses including both excitory and inhibitory. Therefore $g_i$ should be the summarization of all these inputs. The quantity $b_i = c_3 - a_i$ actually means the fatigue of channels. Therefore, postsynaptic potentials of excitory and inhibitory synapses are proportional to the sensitivity and fatigue of membrane channels respectively. And they follow very similar curves (see Fig. 1). In the rest of this paper, synapse is excitory without explicitly indication and motor neuron is called neuron for short. Incidentally, the exponential functions in this model can be easily implemented in physics and biology.

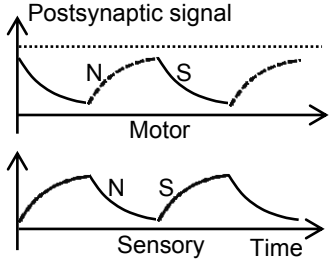

**Fig. 1. Postsynaptic signal of motor neuron and sensory neuron.** Symbol S and N mean cases with and without presynaptic spike respectively. Specially, symbol N in the motor curve can also represent the case when hyperpolarized. And the dotted line means that there is no actual postsynaptic signal. The two curves correspond to motor neuron and sensory neuron respectively, and they are approximately reversal. However, they are caused by different reasons. Specifically, the sensory curve at the bottom is due to the summarization and decay of postsynaptic signal itself. The motor curve however is due to the changes of channels' sensitivity. Therefore it has a specific upper bound. Moreover, both excitory and inhibitory motor neurons share the same curve at the top. And inhibitory neurons are essential for motor learning.

Similar to sensory neuron, S1 actually reflects the spatial summarization of potentials. And the firing frequency f can be viewed as a probabilistic estimate on conditions of features $f_i$ (*10*). However S2 is very different. And physiological evidences support that motor neuron should have a different computational model from sensory neuron. An important difference is that motor neuron can generate dendritic AP (action potential) independent of soma other than EPSP (excitatory postsynaptic potentials) (*8*). As results, there is no temporal summarization of postsynaptic signal as in sensory neuron. In addition, stimulation of high frequency induces LTD (long term depression) in motor neurons (*13*), but it induces LTP (long term potentiation) in sensory neurons (*14*). Since inducing LTP needs the coincidence of soma's BAP (back-propagating AP) and postsynaptic signals (*15*), LTD should be due to lacking BAP. This supports S2 in that membrane channels will become fatigue (see Fig. 1). On the other hand, to induce LTP, BAP should be coincidence with hyperpolarization followed by depolarization (*16*). This supports that membrane channels could recover when hyperpolarized. T-type channels could be the molecular mechanism of N2 (*17, 18*). These channels can be activated at moderate voltages, but require hyperpolarization to firstly remove inactivation so that the channels can open.

Synapse model is as following: S1) LTP: $\frac{dw}{dt} = c_1 s_p h(c_2 - w) > 0$, $\frac{dr}{dt} = c_3 s_p h(1-r) > 0$, and $\frac{dw}{dt} = c_4 w \log(r) < 0$ when $s_p = 0$, where h is the hormone factor, r is the synaptic decay rate, $s_p \approx c_5 f a_i$ according to N2. S2) LTD: $\frac{dw_d}{dt} = -c_6 s_d w_d \ (0 < w_d < 1)$, $\frac{dr_d}{dt} = -c_7 s_d r_d \ (0 < r_d < 1)$, and

$\frac{dw_d}{dt} = -c_8 \log(r_d)(1-w_d)$ when $s_d = 0$, where $r_d$ is the recovery rate of $w_d$, $s_d \approx c_9(c_0 - a_i)$.
LTD here is actually viewed as devaluation of LTP, and the actual synaptic strength should be $ww_d$. BAP or f=0 is the threshold between LTP and LTD. For inhibitory synapses, $s_p \approx c_9 f(c_0 - a_i)$ and $s_d \approx c_9 a_i$. Generally speaking, synapses are alive and self-adaptive: continuous exercises make them thick and tough; without stimulus however they will decay passively with time. From the statistical viewpoint, synaptic strength reflects the confidences of features based on stimulus history. LTD is an adaptive mechanism in essence, similar to fatigue of photoreceptors in the retina.

The model of dendritic competition is similar as in sensory neuron (*10*): C1) $\frac{df_i}{dt} = -p$, $p = c_1(1 - e^{-c_2 \sigma})$, $\sigma = \sum_{j=1, j \neq i}^{n} f_j f_j'$, where p is the total quantity of retrograde messengers, n is the number of neurons connected to the same axonal branch, $f_j$ and $f_j$' are the postsynaptic spike frequencies of dendrite and soma respectively. Specifically, BAP will release retrograde messengers to the presynaptic according to the dendritic spikes frequency and transitorily depress other dendrites connecting to the same axonal branch (*19*). This will lead to the "winner-take-all" competition between postsynaptic dendrites. The dendrite with BAP of the most high frequency will be the unique winner, although it will have been depressed somewhat as well. And winner's dendritic spike frequency should be $f_i = c_1 e^{-c_2 \sigma} + c_3$. This has implications in statistics. Namely, contradictory answers will result in dropping of every answer's reliability. For example, the watch law tells us that it is hard to know the accurate time with many inconsistent clocks. Due to the dendritic competition, a fired axonal branch can send signal to only one postsynaptic neuron. And this will lead to lateral inhibition between neurons sharing common inputs, known as "soft max" or "winner-take-all" in existing theories (*20, 21*). However, our model doesn't need extra "MAX" layers or lateral inhibitory connections. The presynaptic axonal branches themselves actually play similar roles.

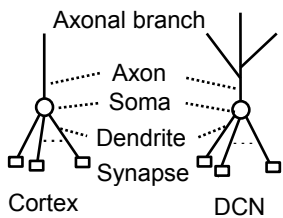

**Fig. 2. Neurons in cerebellar cortex and DCNs.** These two types of motor neurons are responsible for coding proprioception and motor commands respectively. Neurons in DCNs have lower firing threshold. Therefore from the logic viewpoint, they are "OR" gates while neurons in the cerebellum are "AND" gates. Moreover, neurons in DCNs have multiple axonal branches. And dendritic competition only takes effect on single axonal branch independently, due to the limited range of retrograde messagers. PCs are special: they belong to the cortex in general, but they have multiple axonal branches similar to neurons in DCNs.

Suppose free dendritic and axonal branches connect to each other randomly. And synapses grow and decay following the synapse model, neuron model and competition model. Then a neural network and the corresponding learning mechanism will emerge automatically. In general, neurons in the motor system can be divided into two main classes: afferent and efferent (see Fig. 2). In the cerebellum for example, PC (Purkinje cell) is their boundary. In the basal ganglia however, all neurons belong to the latter class. Coding in the afferent end is similar to the cerebral cortex (*10*). In essence the proprioception information including motive direction, amplitude and velocity at any time is an input. Neurons compete and only winners can strengthen their dendritic synapses (*22*). Other synapses decay until broken. Therefore, any input is actually either encoded by a free neuron or merged into the most similar coding neuron. In other words, every input will converge to a single PC through an inward tree structure (see Fig. 3). This PC is actually the coding neuron or "grandmother cell" of the input (*23*), while its revivals can be viewed as background noises. Form this viewpoint, lateral inhibition is a spatial adaptive mechanism, just like LTD being the temporal adaptive mechanism. On the other hand, every PC corresponds to an outward tree structure in the efferent end meanwhile. The inward and outward trees together will compose a sandglass structure (see Fig. 3) (*20*), which actually represents the sensory-motor integration (*1*). Due to the redundancy of freedom degrees (*3*), multiple commands of coding neurons could correspond to the same movement. In the "grandmother cell" coding manner, only one coding neuron can send command at any time. Population coding however will bring difficulties for sensory-motor integration. After all you can't take two conflicting movements at the same time. Instead you have to make choice between them.

As shown in Fig. 4A, Pyramid cells in cerebral cortical motor areas can also be the middle junctions of sandglass structures (*24*), whose efferent end could be the basal ganglia, DCNs or muscles. Different from PC however, Pyramid cells are sensory neurons. Therefore they are more sensitive to continuous inputs. PCs and other motor neurons however are more sensitive to the changing of inputs other than themselves (*18*). These Pyramid neurons are sometimes called "mirror neurons" (*24, 25*), because they will be fired when you see the corresponding action trajectory as well as actually do them. In other words, they can represent both movement desires and corresponding commands. Whether actual actions would be executed is determined by all excitory or inhibitory fibers casting to the corresponding motor units. For example, sometimes you can't help imitating a song, while in other times you can't make a sound in a speech. In the rest of this paper, all these junction neurons are called coding neurons. And when discussing motor learning mechanism, we mainly mean the outward trees in the efferent end including basal ganglia and DCNs (see Fig. 2 and Fig. 3).

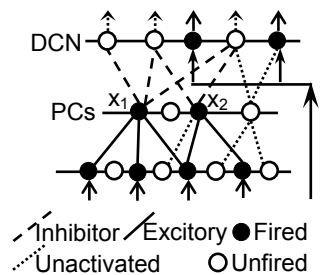

**Fig. 3. Sensory-motor integration and motor learning.** Sensory-motor integration is implemented through sandglass structures with PCs or Pyramid cells as the junctions. There

exists lateral competition between neighbor coding neurons. Due to dendritic competition, neurons fired meanwhile such as $x_1$ and $x_2$ tend to connect the same motor units. This is the conditional reflex in essence, which is the physiological foundation of motor learning.

Every movement corresponds to an efferent command from a coding neuron to motor units, namely an vector: $V= (f_1,f_2,…f_i,…,f_n)$ where $f_i$ is neuron's firing frequency. A complex action corresponds to a sequence of output vectors (*2*), namely $f_i$ changing with time. In the outward tree structure, every neuron has multiple divergent axonal branches, and dendritic competition takes effect in each single axonal branch independently (see Fig. 2). Neurons fired mostly will inhibit neighbors through dendritic competition. As a result of this "rich-get-richer" competition, excitory axonal branches tend to connect to fired neurons, while inhibitory axonal branches tend to connect to hyperpolarized neurons (see Fig. 3). In other words, axonal branches of the same type fired meanwhile tend to connect to the same succeeding neurons. This could be the physiological foundation of conditional reflex, which is the kernel motor learning mechanism. In the conditional reflex experiment for example, since the food and ring often occur meanwhile, ring alone will cause salivating as well. From the computational viewpoint, conditional reflex is a supervised learning in essence (*26*), and inhibitory neurons are essential. As well known in machine learning, both positive and negative samples are needed in supervised learning. According to our model, both CF (climbing fiber) and MF (mossy fiber) should be the teacher signals (*4*) other than error signals (*5*). From the logic viewpoint, neurons in the inward trees are like "AND" gates, while neurons in the outward trees are "OR" gates. The logic functions are mainly determined by firing threshold (see Fig. 2). It can be proved that any logic expression can be represented by such sandglass structures.

A possible question might be how initial sensory-motor pathways form? This question is important for imprinting behavior and language imitating. According to our model, this could be implemented through unsupervised learning (*26*). Random spontaneous actions such as babbling will generate internal and external sensory feedbacks, which will be encoded by PCs and Pyramid neurons. And theses coding neurons tend to connect to those spontaneously fired neurons reversely. As a result, the initial sensory-motor pathways will be built. And then when hearing the pronunciation of a word, the baby would repeat it. Experiments support that babbling is the foundation of language learning (*25*). Any possible action can be formed through unsupervised learning mentioned above. And self-study of skills such as swimming needs repeatedly random tries. However, some actions should be reinforced selectively because are more important for the survival. This could be implemented by reinforcement learning through hormone such as dopamine (*27*). In essence, hormone is meta-mechanism adjusting firing frequency and synaptic growth according to N1 and S1. And this adjustment is determined by evaluations from the instinct. Incidentally, both the supervised learning and reinforcement learning here are "soft" or statistical rather than absolute.

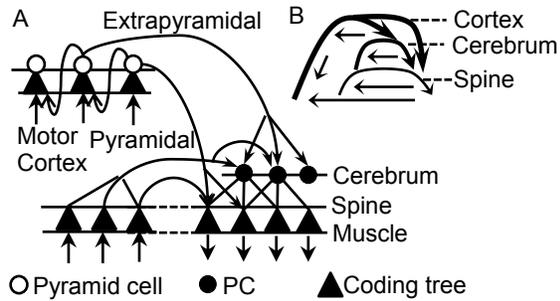

**Fig. 4. Motor system diagram.** In A, sensory-motor integration is implemented through sandglass structures, whose junctions could be Pyramid cells or PCs. The neural network is actually composed of overlapping coding trees. The efferent network is hierarchical and self-similar, and motor learning in every layer following similar mechanism as shown in Fig.2. And lower layers are regulated by high ones. Panel B is the simplification of panel A with extra feedback fibers attached. The main structure is determined by genes according to Crick.

There are controversies about the roles of cerebellum and basal ganglia (*26*). An interesting question is why we need a cerebellum since we have already owned so complex cerebrum. Our model has provided an answer. According to N2, postsynaptic signals decay to zero with input continuing (see Fig. 1). Therefore fierce actions have shorter durations. This can prevent muscle damage caused by continuous fierce contraction. Since single movement is quick and transient, a continuous fierce action actually needs different motor units firing alternately. The strength of a movement is determined by the firing frequencies of motor units in essence. It is adjustable through changing the firing ratio of excitory inputs to inhibitory inputs. And this will make a distinction between quick muscle and slow muscles (*7, 8*). On the other hand, complex skills such as swimming contain many small movements. And the precise intervals between movements are important as well as the strength of every movement (*7*). The interval is actually determined by the length of pathway or circuits. Specifically, since lack of temporal summarization, a motor neuron's launching time is approximately constant for a specific input. Therefore the time of a signal passing through a circuit is proportional to its length. As results, circuits composed of motor neurons completely are punctual and precise. This is important for motive rhythm, ballistic action and keeping balance.

The cerebellum contains such circuits. For example, small circuits constructed by Golgi cells should be important for fine coding. Such small circuit can be viewed as a feedforward control or open loop control in existing theories (*1-3*). Different from the pyramid system, the efferent end of cerebellum is a hierarchical network (*21*). And this is good at fine-grain coding and saving resource (*28*). Moreover, since PCs are inhibitory, the cerebellum is mainly passive and assistant. The circuits in cerebral cortex however aren't punctual due to temporal summarization of postsynaptic signals in sensory neurons (see Fig. 1). The times of signals passing through them will be influenced by signal history. Therefore, the cerebrum isn't as fine or precise as cerebellum. On the other hand, the temporal summarization is actually a kind of instantaneous memory, which could be the foundation of consciousness according to Francis Crick (*29*). Therefore the cerebral cortex is proper for controlling voluntary movements (*30*). In addition, the cerebral cortex contains large circuits for controlling continuous actions (see Fig. 4), known as feedback control and closed loop control (*3*). Since basal ganglia are sensitive to hormones such as dopamine (*26, 27*), they could play the role of hippocampus in sensory memory (*10*). Specifically, the circuits between basal ganglia and the cortex can encode action sequences (see Fig. 4), just like the episodic memory stored in hippocampus. Therefore the basal ganglia could be responsible of procedural learning relating to routine behaviors or "habits". In essence, stuttering and tics-coprolalia syndrome belong to these habits as well. In conclusion, the cerebellum is a global reflex arc or a prompt and mechanical executor in essence (*31*). The cerebrum is an intelligent and rational decision-maker, while the basal ganglia represent evaluations from the instinct.

   It is hard to believe that learning and memory could be simple random processes following statistical laws. The compare between brain and computer is like comparing our bodies with cars. A car is faster than us when running on the man-made roads, but people can move freely in various landforms. A motor is composed of various specialized parts, and it will be paralyzed only if one of them is broken. Our body however contains hundreds of muscles, each of which is composed of similar muscle fibers. From this viewpoint, muscles are simpler than the motor. Similarly, our brain is composed of billions of similar neurons. The most important mechanism should be how neurons interact with each other rather than the circuit details. At present, a focus of neuroscience is drawing the cortical circuits of fibers casting. This might be influenced by computer. However, computer is actually composed of various specialized logic gates similar to the car. And therefore circuits are precise and the details are essential. The computer is actually more complex than the brain, rather than the opposite. Similar to cars as well, computers can defeat people in some specific problems such as chess and launching rocket, while people are better at general problems such as picture recognition and swimming. In conclusion, brain and muscles are statistical, while computers and cars are precise. They are good at different domains.

**Acknowledgments:** This work was supported by the National Natural Science Foundation of China (Grant No. 61170156 and 60933005).